\documentclass
[12pt,showpacs,preprintnumbers,nofootinbib,preprint,amsfonts,amssymb,titlepage,showkeys]{revtex4}%
\usepackage{amsmath}
\usepackage{graphicx}
\usepackage{bm}
\usepackage{amsfonts}
\usepackage{amssymb}
\usepackage{appendix}%
\begin{document}
\preprint{ }
\title{Light Dark Matter and Dark Radiation}
\author{Jae Ho Heo}
\email{jaeheo1@gmail.com}
\author{C.S. Kim}
\email{cskim@yonsei.co.kr}
\affiliation{Department of Physics and IPAP, Yonsei University, Seoul 120-749, Korea }

\begin{abstract}
\noindent Light ($M\leq20$ MeV) dark-matter particles freeze out after
neutrino decoupling. If the dark-matter particle couples to a neutrino or an
electromagnetic plasma, the late time entropy production from dark-matter
annihilation can change the neutrino-to-photon temperature ratio, and equally
the effective number of neutrinos $N_{\text{eff}}$. We study the
non-equilibrium effects of dark-matter annihilation on the $N_{\text{eff}}$
and the effects by using a thermal equilibrium approximation. Both results are
constrained with Planck observations. We demonstrate that the lower bounds of
the dark-matter mass and the possibilities of the existence of additional
radiation particles are more strongly constrained for dark-matter annihilation
process in non-equilibrium.

\end{abstract}

\pacs{95.35.+d, 98.80.Cq }
\keywords{dark matter, dark radiation, effective number of neutrinos}\maketitle

\section{Introduction}

Photons and neutrinos are the lightest particles in the Standard Model (SM)
and give the radiation energy density at late times in the early universe. The
SM neutrino species contributes three degrees of freedom because there are
exactly three neutrino mass eigenstates $\left(  \nu_{1},\nu_{2},\nu
_{3}\right)  $ in combinations of the three flavor eigenstates $\left(
\nu_{e},\nu_{\mu},\nu_{\tau}\right)  $ of the weak interaction. The weak
interactions that keep neutrinos in thermal contact with the electromagnetic
plasma become ineffective around a second after the Big Bang. Neutrinos
decouple at a temperature on the order of $2-3$ MeV before $e^{\pm}$ pairs
annihilate and, thus, do not share in the entropy transfer from $e^{\pm}$
pairs. This causes the neutrino temperature to be less than the photon
temperature later. However, neutrino decoupling was not quite complete when
$e^{+}e^{-}$ annihilation began, so some of the energy and the entropy of
photons could transfer to neutrinos. If the dark radiation density is
parameterized in terms of the number of effective neutrino species
$N_{\text{eff}}$ with the canonical neutrino-to-photon temperature ratio,
$N_{\text{eff}}$ increases to slightly more than the three neutrino species,
leading to $N_{\text{eff}}^{\text{SM}}=3.046$ \cite{Dad82,Gma02}. Because the
number of effective neutrino species $N_{\text{eff}}$ is precisely predicted
in the SM, this can give a robust constraint to any nonstandard physics. For
example, new relativistic particles, such as the light sterile neutrino
\cite{Kna12} or the Goldstone boson \cite{Swe13} which has a decoupling
temperature less than 100 MeV, arise in many extensions of the SM, and their
existence will contribute to the dark radiation energy density. This scenario
is, however, strongly excluded at over the $3\sigma$ level in the latest
Planck analysis \cite{Par15} unless photons or electrons (positrons) are
heated at a later time \cite{Cmh13,Gst13,Cbo13,Kmn14}.

According to a recent analysis of the cosmic microwave background (CMB)
temperature anisotropy by the Planck satellite \cite{Par15}, $N_{\text{eff}}$
was found to be $3.15\pm0.23$ $(1\sigma)$, consistent with the SM prediction.
We should recognize that the Hubble constant $\left(  H_{0}=67.8\pm0.9\text{
kms}^{-1}\text{Mpc}^{-1}\right)  $ inferred by Planck is in tension at about
$2.4\sigma$ with the direct measurement of $H_{0}=73.8\pm2.4$ kms$^{-1}%
$Mpc$^{-1}$ by the HST \cite{Agr11}: larger values of the Hubble constant
prefer larger values of $N_{\text{eff}}$. $N_{\text{eff}}$ was not strongly
excluded to about the $2\sigma$ upper limit in the Planck analysis.
Additionally, $N_{\text{eff}}$ can be inferred from big bang nucleosynthesis
(BBN) considerations \cite{Jya79} at times earlier than recombination because
the theoretical expectations for the primordial abundances of light elements
depend on $N_{\text{eff}}$. Recently, two groups announced different results
for $N_{\text{eff}}$ determined from an analysis of $^{4}$He abundance
measurements in combination with the D abundance \cite{Rco14,Mpe12}. One group
\cite{Eav13} obtained a result consistent with the Planck observation, but the
other \cite{Yiz14} found a larger value of $N_{\text{eff}}\sim3.58$. This
probe does not have the same resolving power as the Planck satellite. We will
use Planck results to constrain the dark radiation in this work.

Recently, weakly interacting massive particles (WIMPs) with sub-GeV masses
\cite{Ewk86,Pds04,Cbo04,Cboe04,Dho04,Kah05}, referred to as light dark matter
(DM), have received some attention because the existence of WIMPs with mass
less than 20 MeV can modify the early universe energy's and entropy densities.
If DM particles couple to the SM particles (neutrinos, or photons and $e^{\pm
}$ pairs) and are sufficiently light ($M\leq20$ MeV) to annihilate after
neutrino decoupling, the primordial plasma will be heated by DM annihilation.
This will affect the neutrino-to-photon temperature ratio and so might be able
to explain possible differences of $N_{\text{eff}}$ from the SM prediction or
to avoid strong experimental constraints on the existence of additional
radiation particles. This scenario was studied in the equilibrium version for
neutrino heating \cite{Cbo12,Gst13,Cbo13} and photon heating
\cite{Cmh13,Gst13,Kmn14,Cbo13}. The equilibrium version is, however, a rough
approximation because DM particles are nonrelativistic at
freeze-out\footnote{We distinguish terminology, \textquotedblleft decoupling"
and \textquotedblleft freeze-out", in this paper. \textquotedblleft
Decoupling" will be used in the case that (DM) particles are completely
non-interacting at some point, and \textquotedblleft freeze-out" is for
chemical decoupling. Notice this does not mean that DM dumps energy in the
neutrino or the electromagnetic plasma instantaneously. The evolving comoving
number density of DM particles has a sizable deviation from its equilibrium
prediction around freeze-out (see Fig. 1).}. If the DM particles are
relativistic at DM decoupling, the equilibrium method must be a good
approximation because they will decouple at equilibrium concentrations. A more
accurate description of the freeze-out process should be considered when DM
particles are nonrelativistic at freeze-out. Here, the Boltzmann equation is
applied to the time-evolution of the DM number in a spatially homogeneous and
isotropic universe. We treat an adiabatic\ expansion of the universe so that
the total entropy stays constant and the second law of thermodynamics can be
applied to the entropy (temperature) evolution of the produced relativistic
particles. Because the DM mass $(M)$ determines the dark radiation energy, it
will be the parameter that we will constrain. We start to study the DM number
evolution for an expanding universe and dark radiation $(N_{\text{eff}})$ in
the equilibrium approximation. Then, we study the out-of-equilibrium
light-particle production, its entropy (temperature) evolution and its effect
on the dark radiation. The possibility of the existence of new light species
(equivalent neutrinos) is also investigated.

\section{Theoretical Details and Numerical Results}

The dark radiation energy density of the universe $\rho_{_{\mathrm{DR}}}$ is
parameterized in terms of the energy density of photons $\rho_{\gamma}$ and
the effective number of neutrinos $N_{\text{eff}}$ with the neutrino-to-photon
temperature ratio of the SM given by%
\begin{equation}
\frac{\rho_{_{\mathrm{DR}}}}{\rho_{\gamma}}=\frac{7}{8}N_{\text{eff}}\left(
\frac{T_{\nu}}{T_{\gamma}}\right)  _{\text{SM}}^{4}.
\end{equation}
The factor $7/8$ is due to the effect of Fermi-Dirac statistics on the energy
density. The exact formula for $N_{\text{eff}}$ depends on the scenario
(model). Because the temperature will change in our scenario, $N_{\text{eff}}$
can be expressed by%
\begin{equation}
N_{\text{eff}}=N_{\text{eff}}^{\text{SM}}\left(  \frac{T_{\nu}}{T_{\gamma}%
}\right)  ^{4}\left(  \frac{T_{\nu}}{T_{\gamma}}\right)  _{\text{SM}}%
^{-4}\text{.}%
\end{equation}
We have assumed that DM particles are nonrelativistic at the time we consider,
for example, the time of DM freeze-out or recombination. The BBN imposes
limits on $N_{\text{eff}}$ at photon temperatures around $1-0.1$ MeV, and any
additional dark radiation particle is unfavorable from BBN considerations
though a small possibility still exists. If the DM particle is relativistic in
the BBN era, it becomes a dark radiation particle. We investigate the effects
of DM annihilation on $N_{\text{eff}}$ for the DM mass range of $0.1-20$ MeV.
The ratio of the neutrino temperature to the photon temperature can be
determined by entropy conservation because the total entropy stays constant in
an adiabatic expansion of the universe. After neutrino decoupling, the
primeval plasma will consist of two decoupled components, the electromagnetic
component and three neutrino ones. The entropies of the neutrino and the
electromagnetic plasmas are separately conserved, and this must serve as an
efficient tool for the study of dark radiation here.

Two independent thermal baths exist after neutrino decoupling, and we will
consider the DM interaction (annihilation) in each thermal bath. Here, the DM
particle always interacts with the plasma in thermal bath $a$, and thermal
bath $b$ is not relevant to the DM interaction, $i.e.$, $a,b=\nu$ or $\gamma$,
but $a\neq b$.

\subsection{Dark-matter Number Evolution}

DM particles are in thermal contact with the rest of the cosmic plasma at high
temperatures, but they will experience the freeze-out at a critical
temperature. In this case, we should consider the Boltzmann equation for the
evolution of DM number. Because the DM interacts with one of the plasmas, we
express the comoving number density about the temperature of the plasma in
thermal bath $b$. This notation is very useful because one plasma can always
be in the thermal equilibrium. If the DM interacts with the plasma in thermal
bath $a$, the evolution equation for the comoving number density $Y$ $(\equiv
n_{DM}/s_{b})$ with respect to the inverse temperature $x_{b}$ $\left(  \equiv
M/T_{b}\right)  $ of the other thermal bath reads
\begin{equation}
\frac{dY}{dx_{b}}=-\frac{\left\langle \sigma v\right\rangle s_{b}}{x_{b}%
H}\left(  Y^{2}-Y_{\text{eq}}^{2}\right)  ,
\end{equation}
where $H$ is the Hubble parameter, $s_{b}$ is the entropy density in thermal
bath $b$ and $Y_{\text{eq}}$ $\left(  =n_{eq}/s_{b}\right)  $ is the
equilibrium number density. This equation is not meaningful when
$Y=Y_{\text{eq}}$. This becomes the usual fluid equation in thermal
equilibrium. We parameterize the annihilation cross section as $\left\langle
\sigma v\right\rangle =\sigma_{0}x_{b}^{-n}$, in which $n=0$ for $s$-wave
annihilation and $n=1$ for $p$-wave annihilation. The above equation can be
reduced to%
\begin{equation}
\frac{dY}{dx_{b}}=-\sqrt{\frac{\pi}{45}}m_{_{\mathrm{PL}}}M\sigma_{0}\left(
\frac{g_{\ast s}^{b}}{\sqrt{g_{\ast}}}\right)  x_{b}^{-n-2}\left(
Y^{2}-Y_{\text{eq}}^{2}\right)  ,
\end{equation}
where $g_{\ast}$ and $g_{\ast s}^{b}$ are the effective relativistic degrees
of freedom for the energy density and the entropy respectively, and
$m_{_{\mathrm{PL}}}$ is the Planck mass.

Unfortunately, Eq. (4) has no analytic solution. Fig. 1 shows the result of
numerical solutions for $s$-wave annihilation into neutrinos (left panel) and
$p$-wave annihilation into photons (right panel). The DM residual annihilation
into photons can distort the CMB spectrum \cite{Dpf12,Llo13}. This effect
excludes DMs with mass less than 10 GeV for $s$-wave annihilation into
photons. The effect is negligible for $p$-wave annihilation, which is velocity
dependent, so this bound can be evaded. For DM annihilation into neutrinos, we
assumed that the same numbers of neutrinos and antineutrinos of each type were
produced. The number of effective relativistic degrees of freedom $g_{\ast
s}^{b}$ is not related to DM or its production. We could take the value on the
SM base. The $g_{\ast}$ is taken as a constant $\overline{g}_{\ast}$ on
average. The curves were made with the proper values of $\sigma_{0}%
\overline{g}_{\ast}^{-1/2},$ in agreement with the current DM relic density
$\left(  Y_{0}\right)  $ for a Dirac fermion, Majorana fermion, complex scalar
and real scalar with a DM mass of 10 MeV. As we expected, the DM number track
the equilibrium abundance at very high temperatures, $x_{b}<1$. The solution
to the Boltzmann equation starts to deviate significantly from the equilibrium
abundance at around $x_{b}\sim10-11$, and the actual DM abundance $Y$ is
different from $Y_{\text{eq}}$ and $Y_{0}$ for a considerable time. Notice
that the equilibrium number densities at high tenperatures (early times) are
different for different particle species. Because the current DM relic density
is not relalted to the particle species, a large number of relativisitic
particles will be produced for the species that has a large equilibrium number
density at early times. This can be a means to distinguish the nature of the
particle if other information about the DM, such as the DM mass, is given (see
Figs. 2 and 3).%

%TCIMACRO{\FRAME{ftbpFU}{16.8635cm}{7.8485cm}{0pt}{\Qcb{Comoving number density
%$Y$ as a function of inverse temperature $x_{\gamma}(=M/T_{\gamma})$ and
%$x_{\nu}(=M/T_{\nu})$ for a $g=1$ real scalar (short dash), $g=2$ Majorana
%(solid), $g=2$ complex scalar (dotted) and $g=4$ Dirac dark matter (long dash)
%with a DM mass of 10 MeV. The $left$ $panel$ is for $s$-wave annihilation into
%neutrinos with $\sigma_{0}\overline{g}_{\ast}^{-1/2}=(2.8-3.0)\times10^{-26}$
%cm$^{3}/$s, and the $right$ $panel$ is for $p$-wave annihilation into photons
%with $\sigma_{0}\overline{g}_{\ast}^{-1/2}=8.0\times10^{-25}$ cm$^{3}/$s$.$
%The horizontal dotted line $(Y_{0})$ represents the current DM relic density,
%and $Y_{\text{eq}}$ indicates the equilibrium number density.}}{}%
%{fig1.eps}{\special{ language "Scientific Word";  type "GRAPHIC";
%width 16.8635cm;  height 7.8485cm;  depth 0pt;  original-width 10.9442in;
%original-height 5.0704in;  cropleft "0";  croptop "1";  cropright "1";
%cropbottom "0";  filename 'fig1.eps';file-properties "XNPEU";}}}%
%BeginExpansion
\begin{figure}
[ptb]
\begin{center}
\includegraphics[
height=7.8485cm,
width=16.8635cm
]%
{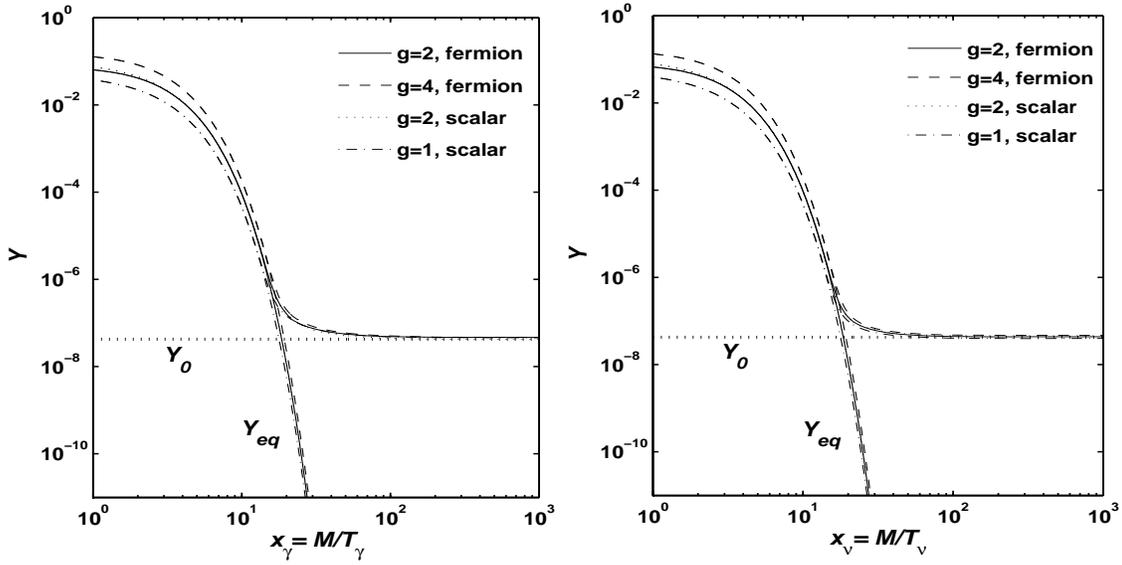}%
\caption{Comoving number density $Y$ as a function of inverse temperature
$x_{\gamma}(=M/T_{\gamma})$ and $x_{\nu}(=M/T_{\nu})$ for a $g=1$ real scalar
(short dash), $g=2$ Majorana (solid), $g=2$ complex scalar (dotted) and $g=4$
Dirac dark matter (long dash) with a DM mass of 10 MeV. The $left$ $panel$ is
for $s$-wave annihilation into neutrinos with $\sigma_{0}\overline{g}_{\ast
}^{-1/2}=(2.8-3.0)\times10^{-26}$ cm$^{3}/$s, and the $right$ $panel$ is for
$p$-wave annihilation into photons with $\sigma_{0}\overline{g}_{\ast}%
^{-1/2}=8.0\times10^{-25}$ cm$^{3}/$s$.$ The horizontal dotted line $(Y_{0})$
represents the current DM relic density, and $Y_{\text{eq}}$ indicates the
equilibrium number density.}%
\end{center}
\end{figure}
%EndExpansion

\subsection{Thermal Equilibrium Approximation}

DM particles are assumed to keep in thermal contact with one of plasmas after
neutrino decoupling and to decouple suddenly at some point. The DM and its
products can be expressed by Fermi-Dirac or Bose-Einstein statistics in this
case. There can be different types of particles in thermal bath $a$, so we
will use the entropy density $s_{a}\equiv\frac{2\pi^{2}}{45}\widetilde
{g}_{\ast s}^{a}T_{a}^{3}=(\rho_{a}+p_{a})/T_{a}$ to define the number of
effective relativistic degrees of freedom\footnote{The mark $``\sim"$ is
placed on top of the symbol of the number of the effective relativistic
degrees of freedom to indicate DM inclusion. If there is no $``\sim"$ mark, DM
is excluded.} $\widetilde{g}_{\ast s}^{a}(T_{a})$ in which $\rho_{a}$ is the
energy density and $p_{a}$ is the pressure. The energy density $\rho_{a}$ and
the pressure $p_{a}$ are expressed by%
\begin{align}
\rho_{a} &  =\sum_{i}\rho_{i}=\sum_{i}\frac{g_{i}}{2\pi^{2}}\int dqq^{2}%
E_{i}\frac{1}{\exp(E_{i}/T_{a})\pm1}\text{,}\\
p_{a} &  =\sum_{i}p_{i}=\sum_{i}\frac{g_{i}}{2\pi^{2}}\int dq\frac{q^{4}%
}{3E_{i}}\frac{1}{\exp(E_{i}/T_{a})\pm1}\text{,}%
\end{align}
where $g_{i}$ is the internal degrees of freedom for the corresponding
particle $i$, $E_{i}=\sqrt{q^{2}+m_{i}^{2}}$ is the energy with mass $m_{i}$,
and the $+(-)$ sign is for fermions (bosons). We set the chemical potentials
to zero. The number of effective relativistic degrees of freedom is given by
\begin{equation}
\widetilde{g}_{\ast s}^{a}\left(  T_{a}\right)  =\frac{45}{2\pi^{2}}%
\frac{\left(  \rho_{a}+p_{a}\right)  }{T_{a}^{4}}.
\end{equation}
Because the entropy in each thermal bath is conserved after neutrino
decoupling, the temperature $T_{a}$ $(T_{b})$ varies as $\widetilde{g}_{\ast
s}^{a-1/3}R^{-1}$ $\left(  g_{\ast s}^{b-1/3}R^{-1}\right)  $, where $R$ is
the scale factor. We can find the temperature ratio at the DM decoupling time
if we know the temperature ratio at a certain time (the time of neutrino
decoupling). The temperature ratio at the time of DM decoupling results in
\begin{equation}
\frac{T_{aD}}{T_{bD}}=\left(  \frac{\widetilde{g}_{\ast s}^{a}\left(  T_{\nu
d}\right)  }{\widetilde{g}_{\ast s}^{a}\left(  T_{aD}\right)  }\right)
^{1/3}\left(  \frac{g_{\ast s}^{b}\left(  T_{bD}\right)  }{g_{\ast s}%
^{b}\left(  T_{\nu d}\right)  }\right)  ^{1/3},
\end{equation}
where $T_{\nu d}$ is the neutrino decoupling temperature at which the photon
and the neutrino temperatures are the same, $T_{aD}$ and $T_{bD}$ are
temperatures\footnote{Notice that one of the DM decoupling temperatures is
determined when the equilibrium DM number is the same as the present-day DM
relic density, $Y_{eq}(T_{bD})=Y_{0}$.} at DM decoupling. This formula can be
approximated to the temperature ratio at late times; $i.e.$, at the times when
the temperatures $T_{a},T_{b}$ are less than the decoupling temperatures.
Because the DM decoupling occurs at $x_{D}\sim18$ as we saw in the subsection
A, there must be almost no DM contribution to the relativistic degrees of
freedom $\widetilde{g}_{\ast s}^{a}\left(  T_{aD}\right)  $. We remove the
mark $``\sim"$. The temperature ratio after DM decoupling is then given by
\begin{equation}
\frac{T_{a}}{T_{b}}\simeq\left(  \frac{\widetilde{g}_{\ast s}^{a}\left(
T_{\nu d}\right)  }{g_{\ast s}^{a}\left(  T_{a}\right)  }\right)
^{1/3}\left(  \frac{g_{\ast s}^{b}\left(  T_{b}\right)  }{g_{\ast s}%
^{b}\left(  T_{\nu d}\right)  }\right)  ^{1/3}.
\end{equation}

We now determine $N_{\text{eff}}$ in each case. If the DM particle interacts
with a neutrino $(a=\nu$ and $b=\gamma)$, the electromagnetic plasma is not
relevant to the DM interaction. We can identify $\left(  g_{\ast s}^{\gamma
}\left(  T_{\gamma}\right)  /g_{\ast s}^{\gamma}\left(  T_{\nu d}\right)
\right)  ^{1/3}$ with $\left(  T_{\nu}/T_{\gamma}\right)  _{\text{SM}}$. We
get the effective number of neutrino species from Eqs. (2) and (9)%
\begin{equation}
N_{\text{eff}}^{\nu}=N_{\text{eff}}^{\text{SM}}\left(  \frac{\widetilde
{g}_{\ast s}^{\nu}\left(  T_{\nu d}\right)  }{g_{\ast s}^{\nu}\left(  T_{\nu
}\right)  }\right)  ^{4/3}\text{.}%
\end{equation}
If the DM particle interacts with an eletromagnetic plasma $(a=\gamma$ and
$b=\nu)$, there is only one species in the neutrino thermal bath, $\nu$. The
effective relativistic degrees of freedom $g_{\ast s}^{\nu}\left(  T_{\nu
}\right)  $ will be the same at any time. Because $\left(  T_{\nu}/T_{\gamma
}\right)  _{\text{SM}}=\left(  g_{\ast s}^{\gamma}\left(  T_{\gamma}\right)
/g_{\ast s}^{\gamma}\left(  T_{\nu d}\right)  \right)  ^{1/3}$, we get the
effective number of neutrino species as%
\begin{equation}
N_{\text{eff}}^{\gamma}=N_{\text{eff}}^{\text{SM}}\left(  \frac{g_{\ast
s}^{\gamma}\left(  T_{\nu d}\right)  }{\widetilde{g}_{\ast s}^{\gamma}\left(
T_{\nu d}\right)  }\right)  ^{4/3}\text{.}%
\end{equation}
%

%TCIMACRO{\FRAME{ftFU}{13.953cm}{8.7733cm}{0pt}{\Qcb{Effective number of
%neutrino degrees of freedom, $N_{\text{eff}}$, as a function of a thermal
%dark-matter mass $M.$ Curves correspond to a $g=1$ self-conjugate scalar
%(short dash), $g=2$ Majorana (solid), $g=2$ complex scalar (dotted) and $g=4$
%Dirac dark matter (long dash). The upper (lower) curves are for the case when
%the dark-matter particles are in thermal equilibrium with neutrinos (electrons
%and photons). The dark horizontal band is the Planck CMB 1$\sigma$ allowed
%range, and the light dark band is the 2$\sigma$ upper allowed range.}}%
%{}{fig2.eps}{\special{ language "Scientific Word";  type "GRAPHIC";
%width 13.953cm;  height 8.7733cm;  depth 0pt;  original-width 9.2907in;
%original-height 5.975in;  cropleft "0";  croptop "1";  cropright "1.0259";
%cropbottom "0";  filename 'fig2.eps';file-properties "XNPEU";}}}%
%BeginExpansion
\begin{figure}
[t]
\begin{center}
\includegraphics[
trim=0.000000in 0.000000in -0.240629in 0.000000in,
height=8.7733cm,
width=13.953cm
]%
{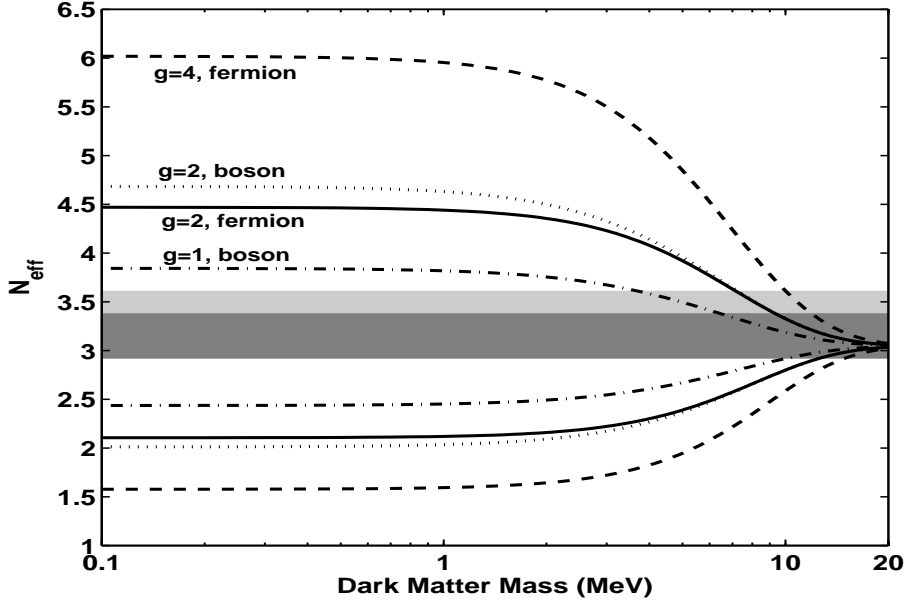}%
\caption{Effective number of neutrino degrees of freedom, $N_{\text{eff}}$, as
a function of a thermal dark-matter mass $M.$ Curves correspond to a $g=1$
self-conjugate scalar (short dash), $g=2$ Majorana (solid), $g=2$ complex
scalar (dotted) and $g=4$ Dirac dark matter (long dash). The upper (lower)
curves are for the case when the dark-matter particles are in thermal
equilibrium with neutrinos (electrons and photons). The dark horizontal band
is the Planck CMB 1$\sigma$ allowed range, and the light dark band is the
2$\sigma$ upper allowed range.}%
\end{center}
\end{figure}
%EndExpansion

\begin{table}[t]
\caption{$1\sigma$ and $2\sigma$ lower limits on the dark-matter mass and
upper limits on the existence of any other dark radiation for dark matter in
thermal equilibrium with neutrinos or electromagnetic plasmas. The mark `$-$'
indicates that the limit is irrelevant. The symbol \textquotedblleft S" stands
for scalar and \textquotedblleft F" for fermion.}%
\begin{ruledtabular}
\begin{tabular}{c|ccc|ccc|cccc}
\multicolumn{1}{c|}{} &\multicolumn{3}{c|}{Neutrino-coupled DM (MeV)}&\multicolumn{3}{c}{EM-coupled DM (MeV)}&\multicolumn{4}{c}{${\Delta}N_\text{eff}$}\\
\hline
$g$&1 (S) & 2 (S,F) & 4 (F) & 1 (S) & 2 (S,F) & 4 (F) & 1 (S) & 2 (S) & 2 (F) & 4 (F) \\
\hline
$1{\sigma}$ & 6.4 & 9.3 & 11.9  & 10.1  & 12.6  & 14.9 & 0.94 & 1.37 & 1.27 & 1.80  \\
$2{\sigma}$ & 3.7 & 7.1  & 10.0  & $-$ & $-$ & $-$ & 1.17 & 1.60 &1.50& 2.03 \\
\end{tabular}
\end{ruledtabular}\end{table}

The curves of Fig. 2 display numerical results for the $N_{\text{eff}}-M$
relation for a Dirac fermion, Majorana fermion, complex scalar and real
scalar. The upper (lower) set of curves are for the case when DM particles are
interacting with neutrinos (electrons and photons). We have implicitly assumed
that neutrinos decouple at $T_{\nu d}$ $\approx2.3$ MeV
\cite{Ken92,Add02,Sha02}. $N_{\text{eff}}$ increases for lighter DM if they
are in equilibrium with neutrinos. Conversely, $N_{\text{eff}}$ decreases for
lighter DM in equilibrium with an electromagnetic plasma. We put a bound on
the DM mass by requiring that $N_{\text{eff}}$ be compatible with the measured
value from Planck \cite{Par15}, and the bounds of the DM masses are listed in
Table I for each species. If there is a significant, but small, density of
additional radiation, the additional radiation density can be explained by
neutrino heating from DM annihilation.\ We should notice that there is still
enough room for the existence of additional radiation particles $\left(
\Delta N_{\text{eff}}\right)  $, such as a sterile neutrino or a Goldstone
boson, with decoupling temperatures less than 100 MeV, when DM
electromagnetically couples to SM particles (EM-coupled DM).

\subsection{Out-of-equilibrium Production}

As we can see in Fig. 1, there is a smooth transition between two regions,
before and after DM freeze-out, and DM particles do not track significantly
the equilibrium from $x_{b}\sim10-11$. The second law of thermodynamics is
applied to the entropy (temperature) evolution of the produced relativistic
particles:%
\begin{equation}
dS_{a}=\frac{dQ}{T_{b}},
\end{equation}
where $dQ=d\left(  R^{3}\rho_{_{\mathrm{DM}}}\right)  $ is the heat added per
comoving volume due to DM annihilation. Because the number of DM particles is
reduced by their annihilation at temperature smaller than DM mass, the energy
density of DM can be described in its nonrelativistic approximation,
$\rho_{_{\mathrm{DM}}}$ $\simeq n_{_{\mathrm{DM}}}M=$ $Ms_{b}Y$. The change in
entropy\footnote{The plasma in the thermal bath $b$ is always in thermal
equilibrium because it is not relevant to the DM interaction, so the entropy
$S_{b}$ is constant.} is given by%
\begin{equation}
dS_{a}=-S_{b}x_{b}dY\longrightarrow\Delta S_{a}=-S_{b}\int_{i}x_{b}dY~,
\end{equation}
where $i$ is an initial point. We consider the initial point at the time of
neutrino decoupling because it is the last point at which neutrinos and
photons are in thermal contact. DM particles with masses less than 20 MeV must
be in thermal equilibrium in thermal bath $a$ at the initial point. Our
observational point is the time of recombination, much later after freeze-out.
After freeze-out (chemical decoupling), DM continues to scatter off
relativistic SM particles untill DM kinetic decoupling, thus thermalizing the
produced particles. Produced electrons (positrons) or photons will be
thermalized quickly due to the electromagnetic interaction. Thermalization of
neutrinos must be slow. Because neutrinos continue to scatter off DM particles
after freeze-out, the produced neutrinos\footnote{A certain number of
neutrinos can remain in non-equilibrium if their scattering strength is not
enough large. We need to consider the detailed Boltzmann equation with the
scattering cross section for this process. Because our work is not concerned
with any specific model, we assume that the produced neutrinos are in the
equilibrium at recombination. The details with scattering are left for a
future study.} can be in the equilibrium at recombination. The change in
entropy is expressed by%

\begin{equation}
\Delta S_{a}=S_{a}-S_{ai}=\frac{2\pi^{2}}{45}\left[  g_{\ast s}^{a}T_{a}%
^{3}R^{3}-\left(  g_{\ast s}^{a}T_{a}^{3}R^{3}\right)  _{i}\right]  \text{.}%
\end{equation}
The temperature ratio is determined by a combination of Eqs. (13) and (14):%
\begin{equation}
\left(  \frac{T_{a}}{T_{b}}\right)  ^{3}=\left(  \frac{g_{\ast si}^{a}%
}{g_{\ast s}^{a}}\right)  \left(  \frac{R_{i}}{R}\right)  ^{3}\left(
\frac{T_{ai}}{T_{b}}\right)  ^{3}-\frac{g_{\ast s}^{b}}{g_{\ast s}^{a}}%
\int_{i}x_{b}dY\text{,}%
\end{equation}
where $T_{ai}$ is, according to Ref. \cite{Cmh13}, very similar to the
neutrino decoupling temperature described in the SM of the absence of DM.
Using the entropy conservation $\left(  g_{\ast s}^{a}R^{3}\sim T^{-3}\right)
$ in the SM, we can approximate the first term of Eq. (15). The temperature
ratio is given by%
\begin{equation}
\left(  \frac{T_{a}}{T_{b}}\right)  ^{3}\simeq\left(  \frac{T_{a}}{T_{b}%
}\right)  _{SM}^{3}-\frac{g_{\ast s}^{b}}{g_{\ast s}^{a}}\left[
x_{b}Y-\left(  x_{b}Y\right)  _{i}-\int_{i}Ydx_{b}\right]  ,
\end{equation}
where we have introduced the integration method by parts, as a convenience,
for numerical computations. The first term on the right-hand side of Eq. (16)
is just the original temperature ratio in radiation, and the second term
represents a contribution from DM annihilation. We can express the temperature
ratio in each case, neutrino-coupled DM $\left(  a=\nu\text{ and }%
b=\gamma\right)  $ and EM-coupled DM $\left(  a=\gamma\text{ and }%
b=\nu\right)  $.
%TCIMACRO{\FRAME{ftbpFU}{13.953cm}{8.7733cm}{0pt}{\Qcb{Same as Fig. 2, but
%contour lines are for the case when radiation particles are produced from
%dark-matter annihilation in the non-equilibrium method (freeze-out mechanism).
%}}{}{fig3.eps}{\special{ language "Scientific Word";  type "GRAPHIC";
%width 13.953cm;  height 8.7733cm;  depth 0pt;  original-width 9.2907in;
%original-height 5.975in;  cropleft "0";  croptop "1";  cropright "1.0257";
%cropbottom "0";  filename 'fig3.eps';file-properties "XNPEU";}}}%
%BeginExpansion
\begin{figure}
[ptb]
\begin{center}
\includegraphics[
trim=0.000000in 0.000000in -0.238771in 0.000000in,
height=8.7733cm,
width=13.953cm
]%
{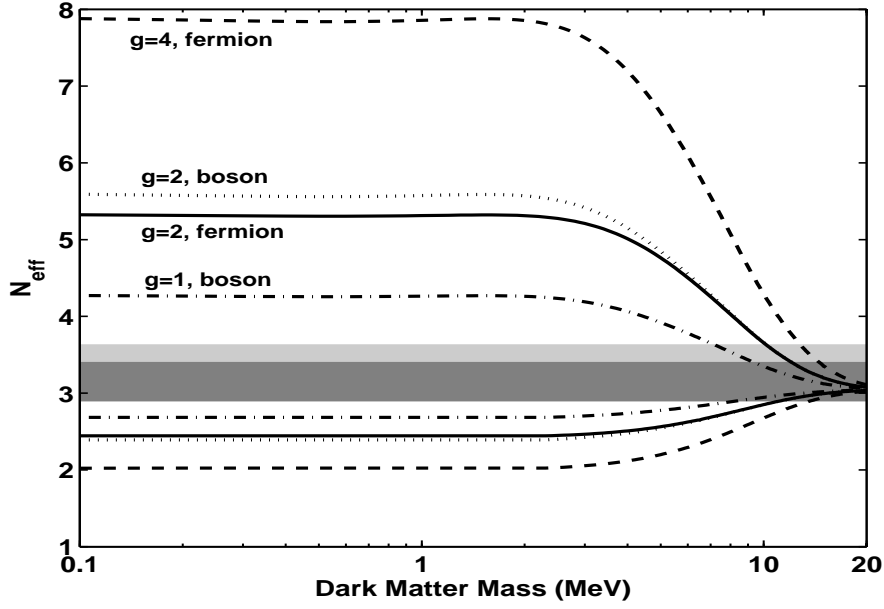}%
\caption{Same as Fig. 2, but contour lines are for the case when radiation
particles are produced from dark-matter annihilation in the non-equilibrium
method (freeze-out mechanism). }%
\end{center}
\end{figure}
%EndExpansion

Fig 3 shows the numerical results of the $N_{\text{eff}}-M$ relation for
neutrino-coupled DM, the upper set of curves, and EM-coupled DM, the lower set
curves. The basic arguments are the same as those in the equilibrium
approximation of subsection B. The bounds on the DM masses are also listed in
Table II, as well as possibilities for the existence of additional radiation
particles $\left(  \Delta N_{\text{eff}}\right)  $. In this DM annihilation
process, the DM mass bounds are more stringent, and the DM effect on the
existence of additional radiation particles is stronger. We interpret this in
the following way: DM particles annihilate more slowly into SM particles to
make a smooth transition. In the radiation-dominant era, $H=$ $\left(
1/R\right)  dR/dt\simeq\sqrt{\left(  8/3\right)  \pi G\rho_{\text{R}}}$, with
the gravitational constant $G=1/m_{\text{PL}}^{2}$ and the radiation energy
density $\rho_{\text{R}}$. The slower annihilation results in a smaller
expansion of the universe and eventually a smaller size of the universe later.
The same number of relativistic particles must be produced from DM
annihilation in the equilibrium and the non-equilibrium processes. The
predicted energy densities at a later time are larger than they are in the
equilibrium process, so DM annihilation effects are larger in the
non-equilibrium process.

\begin{table}[t]
\caption{Same as Table. I, but the values are for the case when radiation
particles are produced from dark-matter annihilation in the non-equilibrium
method (freeze-out mechanism). }%
\begin{ruledtabular}
\begin{tabular}{c|ccc|ccc|cccc}
\multicolumn{1}{c|}{} &\multicolumn{3}{c|}{Neutrino-coupled DM (MeV)}&\multicolumn{3}{c}{EM-coupled DM (MeV)}&\multicolumn{4}{c}{${\Delta}N_\text{eff}$}\\
\hline
$g$&1 (S) & 2 (S,F) & 4 (F) & 1 (S) & 2 (S,F) & 4 (F) & 1 (S) & 2 (S) & 2 (F) & 4 (F) \\
\hline
$1{\sigma}$ & 9.6 & 12.3 & 14.8  & 9.1  & 11.8  & 14.3 & 0.70 & 0.99 & 0.94 & 1.36  \\
$2{\sigma}$ & 7.3 & 10.3  & 12.9  &$-$ &$-$ & $-$ & 0.93 & 1.22 &1.16& 1.56 \\
\end{tabular}
\end{ruledtabular}\end{table}

\section{Conclusions}

Light ($M\leq20$ MeV) dark-matter particles freeze out after neutrino
decoupling. If the dark-matter particle interacts with a neutrino or an
electromagnetic plasma, the late-time entropy production from dark-matter
annihilation can change the neutrino-to-photon temperature ratio, and equally
the effective number of neutrinos $N_{\text{eff}}$. We studied the effects of
dark-matter annihilation on the $N_{\text{eff}}$ by using the thermal
equilibrium approximation and non-equilibrium method (freeze-out mechanism),
and both results were compared with Planck observations. If a significant, but
small, density of additional radiation exists, this can be explained by
neutrino heating from dark-matter annihilation. The effective number of
neutrino species $N_{\text{eff}}$ is reduced for photon heating. In that case,
the existence of additional dark radiation particles can help improve the
agreement with the current observations. The dark-matter particles annihilate
more slowly into SM particles for dark matter annihilation in non-equilibrium.
The slower annihilation results in a smaller expansion rate (eventually a
smaller universe later). Although the same number of relativistic particles
are produced from dark-matter annihilation in the equilibrium approximation
and the non-equilibrium method, the predicted energy densities at a later time
are different. We demonstrated that the lower bounds on the dark-matter mass
and the possibilities of the existence of additional radiation particles are
more strongly constrained for dark-matter annihilation process in non-equilibrium.

\begin{acknowledgments}
The work is supported by a National Research Foundation's Korea (NRF) grant
funded by Korea government of the Ministry of Education, Science and
Technology (MEST) (Grant No. 2011-0017430 and Grant No. 2011-0020333).
\end{acknowledgments}

\end{document}